\documentclass{aims}
\usepackage{amsmath,amsfonts,amssymb}

\def\DOI{xxx XXX}

\def\a{\alpha}
\def\b{\beta}
\def\ga{\gamma}
\def\de{\delta}   

\def\vphi{\varphi}
\def\la{\lambda}

\def\s{\sigma}

\def\om{\omega}

\def\vphi{\varphi}

\def\L{\mathcal{L}}

\def\O{{\cal O}}

\def\R{{\bf R}}

\def\T{{\rm T}}

\def\Ga{\Gamma}
\def\De{\Delta}
\def\La{\Lambda}

\def\pa{\partial}

\def\d{{\rm d}}       

\def\o+{\oplus}
\def\xd{{\dot x}}

\def\ss{\subset}

\def\ker{{\rm Ker}}

\def\<{\langle}
\def\>{\rangle}

\def\({\left(}
\def\){\right)}
\def\[{\left[}
\def\]{\right]}
\def\=#1{\bar #1}
\def\~#1{\widetilde #1}
\def\wt#1{\widetilde #1}
\def\.#1{\dot #1}
\def\^#1{\widehat #1}

\def\"#1{\ddot #1}

\def\interno{\hskip 2pt \vbox{\hbox{\vbox to .18
truecm{\vfill\hbox to .25 truecm
{\hfill\hfill}\vfill}\vrule}\hrule}\hskip 2 pt}

\def\eeq{\end{equation}}
\def\beq{\begin{equation}}

\def\beql#1{\begin{equation} \label{#1}}

\def\eqref#1{(\ref{#1})}

\def\symmref{AVL,CGbook,KrV,Olv1,Olv2,Ste}
\def\MuRaltri{MuRom3,MuRom4,MuRom4b,MuRom5,MuRom6,MuRom7,MuRom11,MuRom11b,MuRom12,MuRom14}

\def\mapright#1{\smash{\mathop{\longrightarrow}\limits^{#1}}}
\def\mapdown#1{\Big\downarrow\rlap{$\vcenter{\hbox{$\scriptstyle#1$}}$}}

\def\interno{\hskip 2pt \vbox{\hbox{\vbox to .18
truecm{\vfill\hbox to .25 truecm
{\hfill\hfill}\vfill}\vrule}\hrule}\hskip 2 pt}

\begin{document}

\title{On the geometry of twisted prolongations, \\ and dynamical systems}

\author{G. Gaeta}
\address{\it Dipartimento di Matematica, Universit\`a degli Studi di
Milano, via Saldini 50, 20133 Milano (Italy); e-mail: {\tt giuseppe.gaeta@unimi.it}; ORCID: {\tt 0000-0003-3310-3455}}

\maketitle

\begin{abstract} \noindent
I give a short review of the theory of twisted symmetries of
differential equations, emphasizing geometrical aspects. Some open
problems are also mentioned.
\end{abstract}

{} \hfill {\it Dedicated to Juergen Scheurle on the occasion of
his retirement}


\section{Introduction}

Sophus Lie created what is nowadays known as the theory of Lie
groups and algebras first and foremost to study (nonlinear)
differential equations. The theory has then been extended in
several directions, in particular generalizing the set of admitted
vector fields. On the other hand, it remained clear that once we
have defined how the basic (independent and dependent) variables
are acted upon by our transformations, the action on derivatives
is given -- by a natural action, known in geometrical terms as the
\emph{prolongation} operation.

More recently, it has been realized that one can also deform the
action on derivatives, i.e. deform the prolongation operation (in
this case one usually speaks of ``twisted prolongation'' and
``twisted symmetry''), and still obtain useful concepts and
results -- where useful is meant in the sense of ``useful to get
solutions of the equations under study'', beside the abstract
geometrical interest.

It happens that in these cases the deformation is assigned at the
level of first derivatives, while deformations on higher
derivative sees no different action. This means that -- as is
often the case in symmetry theory of differential equations -- in
the case of first order equations, even more so for Dynamical
Systems, one has ``too much freedom'' (a standard euphemism to
mean there is no algorithmic way to proceed). Despite this fact,
the theory can also be used in the context of dynamical systems (a
special attention in this direction was paid in the development of
$\s$-symmetries, see below).

In this paper I will review the theory of twisted symmetries,
paying special attention to geometrical aspects -- in particular
to the connection between the usual Lie reduction and
Lie-Frobenius one -- and to results which can be applied in the
realm of ODEs and Dynamical Systems. The Bibliography will provide
the interested reader with indications on how to go beyond these
short notes.

\section{Standard symmetries of differential equations}

I will consider differential equations\footnote{For the moment,
ODEs or PDEs will not make a difference, and differential
equations, are always possibly vector ones, i.e. systems;
similarly, functions are always possibly vector ones -- albeit in
some cases I will use vector indices explicitly to avoid possible
confusion.}  with independent variables $x^i$ ($i=1,...,n$) and
dependent variables $u^a$ ($a = 1,..., m$); partial derivatives
will be denoted by $u^a_J$, where $J$ is a multi-index $J = \{ j_1
, ... , j_n \}$ of order $|J| = j_1 + ... + j_n$ and \beq u^a_J \
= \ \frac{\pa^{|J|} u^a}{\pa x_1^{j_1} ... \pa x_n^{j_n} } \eeq
(here and somewhere in the following we moved the vector index of
the $x$ for typographical convenience). We denote by $u_{(k)}$ the
set of all partial derivatives of order $k$, and by $u_{[n]}$ the
set of all partial derivatives of order $k \le n$.

\subsection{Geometrical description of differential equations and solutions}

The $x$ are local coordinates in a manifold $B$, while $u$ are
local coordinates in a manifold $U$; we consider the phase
manifold $M = B \times U$, which has a natural structure of bundle
$(M,\pi_0,B)$ over $B$ with fiber $U$.

As well known \cite{\symmref} we can associate to $(M,\pi_0,B)$
its Jet bundles $J^k M$ of any order; these have a structure of
fiber bundle $(J^k M, \pi_k , B)$ over $B$ (with projection
$\pi_k$) but also of fiber bundle over those of lower order (with
projection $\chi_{kq}$, for $q < k$), i.e. $(J^k M , \chi_{kq} ,
J^q M)$ with $\pi_k = \pi_q \circ \chi_{kq}$. Natural local
coordinates in $J^k M$ are provided by $\(x,u, u_{(1)} , ... ,
u_{(k)} \)$.

We also recall that the Jet bundle is equipped with a
\emph{contact structure} $\Omega$ \cite{ArnGM,God,Olv2,Sha,Stern};
this can be encoded in the \emph{contact forms}\footnote{Here and
in the following we adhere to Einstein summation convention over
repeated indices (and multi-indices).} \beql{eq:omega} \om^a_J \
:= \ \d u^a_J \ - \ u^a_{J,i} \, \d x^i \ . \eeq

Functions $u = f(x)$ are naturally identified with sections in $M$
(elements of $\Sigma (M)$); the function $u=f(x)$ corresponds to
the section \beq \ga_f \ = \ \{ (x,u) \in M \ : \ u = f(x) \} \ . \eeq
Note that in this way we have established a correspondence between
an \emph{analytical} object (the function) and a
\emph{geometrical} one (the section).

A section $\ga_f \in \Sigma (M)$ identifies naturally a section
$\ga_f^{(k)}$ in $\Sigma (J^k M)$, with of course
\beq \ga_f^{(k)} \ = \ \{ (x,u_{[k]}) \in J^k M \ : \
u^a_J = f^a_J (x)  \ \ \forall J : |J| \le k \} \ . \eeq

Given a differential equation $\Delta$ of order $k$, written in
local coordinates as \beql{eq:Delta} F^i \( x,u,...,u_{(k)} \) \ =
\ 0 \ , \eeq we consider its \emph{solution manifold} $S_\Delta
\ss J^k M$, \beq S_{\De} \ = \ \{ (x,u_{(1)} , ... , u_{(k)}) \, :
\ F \( x,u_{[k]}\) = 0 \} \ . \eeq This is just the set of points
in $J^k M$ where the relation described by $\Delta$ between
independent, dependent variables and derivatives is satisfied; but
now we have again transformed an \emph{analytic} object (the
differential equation) into a \emph{geometric} one.

The same can be done for the concept of solutions to $\De$: a
function $u = f (x)$ identifies, as mentioned above, a section
$\ga_f$ in $(M,\pi_0,B)$, and this in turn identifies a section
$\ga_f^{(k)}$ in $(J^k M,\pi_k , B)$, which is just the set of
points $(x,u_{[k]})$ with $u^a = f^a(x)$ and $u^a_J = f^a_J (x)$.
Now $u = f(x)$ is a solution to $\De$ if and only if $\ga_f^{(k)}
\ss S_\De$.

\subsection{Vector fields and prolongations}

Let us now consider a vector field $X$ in $M$ and its
\emph{prolongation} to $J^k M$. In local coordinates, we write
\beq X \ = \ \sum_i \xi^i (x,u) \,  \frac{\pa}{\pa x^i} \ + \
\sum_a \vphi^a (x,u) \, \frac{\pa }{\pa u^a} \ = \ \xi^i \, \pa_i
\ + \ \vphi^a \, \pa_a \ ; \eeq the prolongation is then described
--in the same local coordinates -- by \beq X^{(k)} \ = \ X \ + \
\sum_a \sum_{|J|=1}^k \psi^a_J \, \frac{\pa}{\pa u^a_J} \ := \ X \
+ \ \psi^a_J \, \pa_a^J \ , \eeq where we introduced the shorthand
notations
\beq \pa_i \ := \ \frac{\pa}{\pa x^i} \ , \ \ \pa_a \ := \
\frac{\pa}{\pa u^a} \ ; \ \ \pa_a^J \ := \ \frac{\pa}{\pa u^a_J} \ . \eeq
We will also write $\psi^a_0 = \vphi^a$.

The coefficient $\psi^a_J$ of the Jet components, i.e. of the
components making the prolongation of $X$, are computed by the
\emph{prolongation formula}, which is more conveniently expressed
in recursive form: \beql{eq:prol} \psi^a_{J,i} \ = \ D_i \psi^a_J
\ - \ u^a_{J,k} \ D_i \xi^k \ . \eeq Here and in the following
$\wt{J} = \{ J,i \}$ is the multi-index with components $\wt{j}_m
= j_m + \delta_{m,i}$.

It is maybe worth recalling that this is easily obtained in
analytic terms (we assume the reader to be familiar with this
derivation, which is however provided in \cite{Gtwist1,Olv1,Olv2}), but the
prolongation of a vector field can also be defined geometrically.

The following Lemmas are well known; see e.g.
\cite{Gtwist1,Olv1,Olv2} for proofs and details.

\medskip\noindent
{\bf Lemma 1.} {\it The prolonged vector field $X^{(n)}$ is the
unique vector field in $J^n M$ which: $(i)$ is projectable to each
$J^k M$ for $0 \le k \le n$; $(ii)$ coincides with $X$ when
restricted to $M$; $(iii)$ preserves the contact structure on $J^n
M$.}

\medskip\noindent
{\bf Lemma 2.} {\it The prolongation of the commutator of two
vector fields is the commutator of their prolongations; in other
words, \beq [X,Y] = Z \ \Leftrightarrow \ \[ X^{(n)} , Y^{(n)} \] \
= \ Z^{(n)} \ . \eeq}
\bigskip

We will now consider \emph{differential invariants} (DIs) for a
vector field $X$; these are invariants for the action of the
prolongation of $X$ in $J^k M$. If a differential invariant
$\zeta$ depends only on variables belonging to $J^k M$ (that is,
no dependence on derivatives of order higher than $k$, and
effective dependence on at least one derivative of order $k$), we
say it is a DI of order $k$. DIs of order zero are ordinary
invariants for the $X$ action in $M$.

\medskip\noindent
{\bf Lemma 3.} {\it Let $\eta : M \to \R$ be a differential
invariant of order zero and $\zeta : J^k M \to \R$ a differential
invariant of order $k$ for $X^{(n)}$ ($n > k$). Then $\chi = (D_i
\zeta / D_i \eta): J^{k+1} M \to \R$ is a differential invariant
of order $k+1$ for $X^{(n)}$.}

\medskip\noindent
{\bf Remark 1.}
Lemma 2 is also formulated saying that prolongation
preserves Lie algebra structures.

\medskip\noindent
{\bf Remark 2.} The property stated in Lemma 3 is also known as
``invariant by differentiation property'' (IBDP); if we start with
a set of invariants of order 0 and 1, we can generate differential
invariants of all orders. In the case of ODEs, if we start with a
complete set of DIs of order zero and one, we can generate in this
way a complete set of DIs of any order, basically because if $U$
is $q$-dimensional, we have $k \cdot q$ DIs of order $k$ (these
includes those of lower orders), as follows at once from $J^k M$
being of dimension $d_k = (k +1) q + 1$. The situation is
different for PDEs, as the dimension of $J^k M$ grows
combinatorially; see e.g. the discussion in \cite{Olv1}.

\subsection{Lie-point symmetries}

A (Lie-point\footnote{More general classes of symmetry can be (and
indeed are, in the literature) considered, but we will only
consider these.}) symmetry, or more precisely a Lie-point symmetry
generator, is a vector field $X$ on $M$ such that its
\emph{prolongation} $X^{(k)}$ to $J^k M$ is tangent to $S_\De$.
For a given $\Delta$ in the form \eqref{eq:Delta}, this condition
is expressed in local coordinates as \beql{eq:deteq} \[ X^{(k)} \(
F^i \) \]_{S_\De} \ = \ 0 \ . \eeq In these equations, also called
\emph{determining equations}, the $F$ are given and one looks for
$\xi$, $\vphi$ (i.e. components of the vector field $X$)
satisfying them. As the components $\psi^a_J$ of $X^{(k)}$ along
$u_{J}$ are given in terms of $\xi,\vphi$ and their derivatives,
all dependencies of $u^a_J$ (with $|J| \not= 0$) are fully
explicit, and hence \eqref{eq:deteq} decouple into a system of
simpler equations, one for each monomial in the $u^a_J$; each of
these is a \emph{linear} PDE for the $\xi$ and $\vphi$, and they
can be solved algorithmically -- usually with the help of a
symbolic manipulation program, as the dimension of the system can
be quite large. This fails in the case of \emph{first order}
equations.

\medskip\noindent
{\bf Remark 3.} The symmetry relation requires the vector field to
be tangent to the manifold representing the equation; this means
that only \emph{integral curves} of vector fields are relevant,
and not the speed on these \cite{CGWs1,CGWs3,Gtwist2,PuS}.

\medskip\noindent
{\bf Remark 4.} A generic equation will have \emph{no} symmetries;
symmetry is a non-generic property -- albeit it may become generic
in a given class of equations: e.g., equations for isolated
physical systems are invariant under time and space translations,
and space rotations; as well known, conservation of Energy,
Momentum and Angular Momentum is related to these invariances via
Noether theorem \cite{Arn,YKS,Olv1}.

\medskip\noindent
{\bf Remark 5.} There can be vector fields $X$ such that
\eqref{eq:deteq} is satisfied without the restriction to $S_\De$,
i.e. such that $X^{(k)} \( F \) = 0$; in this case one speaks of
\emph{strong} (Lie-point) symmetries. The relation between strong
and standard symmetries was clarified by Carinena, Del Olmo and
Winternitz \cite{CDW} (CDW theorem); roughly speaking -- and up to
some cohomological considerations -- if a differential equation
$\De$ admits $X$ as a symmetry, there is always a differential
equation $\wt{\De}$ which admits $X$ as a \emph{strong} symmetry
and such that $\De$ and $\wt{\De}$ have the same set of solutions.

\medskip\noindent
{\bf Remark 6.} In a nineteenth-century language, the advantage of
knowing symmetries of a differential equation is that its
analysis, and the search for its solutions, are much easier if one
uses the ``right'' coordinates, i.e. \emph{symmetry-adapted}
coordinates -- pretty much as analyzing rotationally invariant
problems is easier using spherical coordinates.

\subsection{Symmetry of ODEs}
\label{sec:sode}

The use of symmetry for ODEs is quite simple; let us focus for
simplicity (and for ease of comparison in the case of
$\la$-symmetries to be considered below) on $\De$ a \emph{scalar}
ODE of order $N>1$, \beq F \( x,,u,...,u_{(N)} \) \ = \ 0 \ . \eeq
Suppose we were able to determine a Lie-point symmetry $X = \xi
\pa_x + \vphi \pa_u$ for it, and say it is a strong symmetry (if
not we can use the CDW theorem and consider the equivalent
equation $\wt{\De}$, see above); outside singular points of $X$,
we can pass to coordinates $(y,v)$ such that $X = \pa_v$ (flow box
theorem \cite{ArnODE,ArnGM}); but if in these coordinates $X$ is
written in this way, its prolongation will be $X^{(N)} = \pa_v$,
as follows immediately from \eqref{eq:prol}.

The equation $\De$ will be written in the new coordinates in some
different way, i.e. $\De$ now reads as \beq G \( y,v,...,v_{(N)}
\) \ = \ 0 \ ; \eeq however the fact that it is invariant under
$X^{(N)}$ \emph{and} the peculiar form of $X^{(N)}$ in these
coordinates imply that $G$ does not depend on $v$, i.e. we
actually have \beql{eq:red0} G \( y,v_{(1)},...,v_{(N)} \) \ = \ 0
\ . \eeq It now suffices to make a new change of variables,
introducing $ w := v_y$, to have a differential equation of lower
order, \beql{eq:redODE} H (y, w_{[N-1]}) \ \equiv \ G \(
y,w,...,w_{(N-1)} \) \ = \ 0 \ . \eeq The procedure can be
iterated if this has some further symmetry (see also the Remarks
below). In this way we obtained a (symmetry) \emph{reduction} of
our ODE.

If we are able to solve \eqref{eq:redODE}, say to determine a
solution $w = g (y)$, we can reconstruct a solution $v = v(y)$ to
\eqref{eq:red0} simply by an integral -- in this context one
speaks of a \emph{quadrature} -- i.e. by
\beq v(y) \ = \ \int w(y) \ d y \ . \eeq
Inverting the original change of coordinates we obtain a function
$u = u(x)$, which is a solution to the original equation.

Note that our general notation is redundant for ODEs; in this case
all derivatives are w.r.t. a single variable $x$, and we can
accordingly just write $u^a_{(k)} = d^k u^a / d x^k$, and
similarly $\psi^a_{(k)}$ for the coefficient of $d/du^a_{(k)}$ in
$X^{(m)}$. The prolongation formula \eqref{eq:prol} takes then the
simpler form \beql{eq:prolODE} \psi^a_{(k+1)} \ = \ D_x
\psi^a_{(k)} \ - \ u^a_{(k+1)} \, D_x \xi \ . \eeq

\medskip\noindent
{\bf Remark 7.} Changing variables from $(x,u)$ to $(y,v)$ also means
changing the contact forms from $\om^a_J = \d u^a_J - u^a_{J,i} \d
x^i$ to $\eta^a_J = \d v^a_J - v^a_{J,i} \d y^j$.

\medskip\noindent
{\bf Remark 8.} If the equation has several symmetries, i.e. not only the
one we are using for reduction but some other ones as well, it is
not guaranteed that these will still be present after the
reduction. In general, one can fully use only a (maximal)
\emph{solvable} subalgebra of the symmetry algebra of the
equation, and this provided the generators are used for reduction
in the ``right'' order; see e.g. \cite{\symmref}.

\medskip\noindent
{\bf Remark 9.} On the other hand, the reduced equation could have
symmetries which were not present in the original equation. These
symmetries appearing upon reduction can be ``predicted'', and the
features behind their appearance go under the name of ``solvable
structures''; see e.g. \cite{BP,BH,CFM1,CFM2,HaA,ShP} for details.

\medskip\noindent
{\bf Remark 10.} The possibility of effectively operating symmetry reductions
as sketchily described above depends on the ``invariants by
differentiation property''; we refer again to standard texts
\cite{\symmref} for details.

\subsection{Symmetry of PDEs}

The use of symmetries in the analysis of PDEs is rather different;
actually even the \emph{aim} of using symmetry is different. In
fact, for ODEs one can hope of determining the most general
solution, or at least (as we have seen above) to determine a
reduced equation whose solutions are in correspondence -- via a
quadrature -- to solutions to the original equation.

For nonlinear PDEs looking for the general solution is in general
(i.e. except for integrable equations) a hopeless task, and one
should instead aim at determining at least some solutions. Again
the parallel with the familiar case of rotational symmetries makes
things quite clear: one looks first for \emph{symmetric
solutions}, and such solutions can be determined by solving a
(usually) simpler equation, i.e. one depending on fewer variables.
E.g., rotationally invariant solutions depend just on the radial
coordinate $r$, and hence are determined by an ODE. In the case of
a nonlinear equation this will in general be a nonlinear ODE, and
its solution can still be rather hard, but we definitely face a
simpler problems than the original one -- and correspondingly if
we completely solve it, we have only a partial solution to the
original one.

Thus, while in the ODE case we were looking for new coordinates in
which the symmetry vector field $X$ was along one of the
\emph{dependent} coordinates, in the PDE case we want new
coordinates in which the symmetry vector field $X$ (or vector
fields $X_i$) is (are) along one (several) of the
\emph{independent} coordinates $y^i$. We will then look for
solutions $v = f (y)$ which are invariant under the $X_i$
($i=1,...,r$), i.e. which do not depend on the $(y^1,...,y^r)$;
correspondingly we will have to solve a PDE for $v$ being a
function of the $(y^{r+1},...,y^n)$ variables, i.e. in less
independent variables than the original one.

\medskip\noindent
{\bf Remark 11.} From the geometric point of view, an
$X$-invariant solution $u = f(x)$ is a section $\ga_f \in \Sigma
(M,\pi_0,B)$ such that $\ga_f^{(n)} \in S_\De$ (which ensures it
is a solution) and also such that $X (\ga_f ) = 0$, which of
course also implies $X^{(n)} ( \ga_f^{(n)} ) = 0$. If we consider
a different vector field $\wt{X}$ such that $\wt{X} = \mu X$ on
$\ker (X)$ (here $\mu$ is a smooth function on $M$), such
solutions will also be $\wt{X}$-invariant.

\section{Simple twisted symmetries}

In recent years, starting from the seminal work of Muriel and
Romero in 2001 \cite{MuRom1,MuRom2} (see also \cite{\MuRaltri}), several
kinds of \emph{twisted symmetries} have been considered in the
literature \cite{Gtwist1,Gtwist2}.

The name originated from the fact here one considers a Lie-point
vector field $X$ in $M$, but the prolongation operation is
deformed in a way which depends on an auxiliary object. In
different realizations this can be a scalar function
($\lambda$-symmetries \cite{MuRom1,MuRom2}), a matrix-valued one
form satisfying the horizontal Maurer-Cartan equations -- i.e. a
set of matrices satisfying a compatibility condition
($\mu$-symmetries \cite{CGMor}), or a matrix acting in an auxiliary space
($\sigma$-symmetries \cite{CGWs1}).\footnote{It should be said that
actual ``twisting'' only occurs in the latter cases, not for
$\lambda$-symmetries, but I find it convenient to use this
collective name \cite{Gtwist1,Gtwist2}.}

It should also be stressed that twisted symmetries are more easily
used for \emph{higher order} differential equations (ordinary or
partial), while the case of first order equations is in some sense
degenerate from this point of view, and presents several
additional problems.

\subsection{$\la$-symmetries}
\label{sec:lambda}

The first type of twisted symmetries to be introduced was
$\la$-symmetries (the name $C^\infty$ symmetries also appears in
the literature). These (originally) considered scalar ODEs of any
order, and the name refers to the auxiliary $C^\infty$ function
$\la (t,x,\xd)$ defining the twisted prolongation, which in this
case is called $\la$-prolongation. In fact, this is recursively
defined as
\begin{eqnarray} \psi^a_{(k+1)} &=& D_x \psi^a_{(k)} \ - \
u^a_{(k+1)} \ D_x \, \xi \ + \ \la \, \( \psi^a_{(k)} \ - \
u^a_{(k)} \, \xi \) \nonumber \\ &=& (D_x \, + \, \la )
\psi^a_{(k)} \ - \ u^a_{(k+1)} \ (D_x \, + \, \la ) \, \xi \ .
\label{eq:laprol}  \end{eqnarray} We will denote the
$\la$-prolongation of order $k$ of the vector field $X$ in $M$ as
$X^{(k)}_{\la}$.

The vector field $X$ in $M$ is said to be a \emph{$\la$-symmetry}
of the equation $\De$ (of order $k$) if \beq X^{(k)}_\la \ : \ S_\De
\ \to \T \, S_\De \ . \eeq Note that in general a vector field is a
$\la$-symmetry of a given equation \emph{only} for a specific
choice of the function $\la$.

\medskip\noindent
{\bf Lemma 4.} {In general, the commutator of the
$\la$-prolongations of two vector fields $X,Y$ in $M$ is
\emph{not} the $\la$-prolongation of their commutator, i.e. if $Z
= [X,Y]$ then (in general, for $\la \not= 0$)
\beq \[ X^{(n)}_\la , Y^{(n)}_\la \] \ \not= \ Z^{(n)}_\la \ . \eeq}

\smallskip\noindent
{\bf Proof.} Consider e.g. $X = x \pa_u$, $Y = u \pa_u$; in this
case $Z = [X,Y] = x \pa_u$, and $\delta := [ X^{(1)}_\la ,
Y^{(1)}_\la ] - Z^{(1)}_\la  = x  \la + ( u  -  x u_x )
\la_{u_x}$.

\medskip\noindent
{\bf Lemma 5.} {\it The IBDP holds for $\la$-prolonged vector fields.}

\smallskip\noindent
{\bf Proof.} See e.g. \cite{MuRom1,MuRom2}, or \cite{Gtwist1}.

\medskip\noindent
{\bf Remark 12.} Lemma 5 makes $\la$-symmetries ``as useful as
standardly prolonged ones'', as we will see below in Section
\ref{sec:use}.

\medskip\noindent
{\bf Remark 13.} It was pointed out by Pucci and Saccomandi
\cite{PuS} that $\la$-prolonged vector fields can be characterized
as the \emph{only} vector fields in $J^k M$ with the property that
their integral lines are the same as the integral lines of some
vector field which is the standard prolongation of some vector
field in $M$. This remark was fully understood only some time
after their paper, and was the basis for many of the following
developments, discussed below.

\subsection{$\mu$-symmetries}
\label{sec:mu}

The $\la$-prolongation is specifically designed to deal with ODEs
(or systems thereof); a generalization of it aiming at tackling
PDEs (or systems thereof) is the $\mu$-prolongation. This can of
course also be applied to ODEs and Dynamical Systems.

\subsubsection{PDEs}

Now the relevant object is not a single matrix, but an array of
matrices $\La_i$, one for each independent variable. These are
better encoded as a ($GL(n,\R)$-valued) \emph{horizontal one-form}
\beql{eq:mu} \mu \ = \ \La_i (x,u,u_x) \, \d x^i \ . \eeq The
matrices $\La_i$ should satisfy a compatibility condition, i.e.
\beql{eq:MCH} D_i \, \La_j \ - \ D_j \, \La_i \ + \ \[ \La_i ,
\La_j \] \ = \ 0 \ ; \eeq this is immediately recognized as the
\emph{horizontal Maurer-Cartan equation},or equivalently as a
\emph{zero-curvature condition} for the connection on $\T U$
identified by \beql{eq:nabla} \nabla_i \ = \ D_i \ + \ \Lambda_i \
. \eeq

If $\mu$ satisfies \eqref{eq:MCH}, we can define
$\mu$-prolongations in terms of a modified prolongation formula,
called of course \emph{$\mu$-prolongation formula} (and which
represents now an actual twisting of the familiar prolongation
operation): \begin{eqnarray} \psi^a_{J,i} &=& D_i \psi^a_{J} \ -
\ u^a_{J,k} \ D_i \, \xi^k \ + \ (\La_i)^a_{\ b} \, \( \psi^b_{J}
\ - \ u^b_{J,k} \, \xi^k \) \nonumber \\ &=& (D_i \, I \, + \, \La_i )^a_{\
b} \, \psi^b_{J} \ - \ u^b_{J,k} \ (D_i \, I \, + \, \La_i )^a_{\
b} \, \xi^k \ . \label{eq:muprol} \end{eqnarray}

We will denote the $\mu$ prolongation (of order $k$) of the vector
field $X$ in $M$ as $X^{(k)}_\mu$. The vector field $X$ in $M$ is
said to be a \emph{$\mu$-symmetry} of the equation $\De$ (of order
$k$) if \beq X^{(k)}_\mu \ : \ S_\De \ \to \T \, S_\De \ . \eeq Note
that in general a vector field is a $\mu$-symmetry of a given
equation \emph{only} for a specific choice of the one-form $\mu$.

\medskip\noindent
{\bf Remark 14.} In $\la$-prolongations the prolongation operation
is modified, but it acts separately on the different vectorial
components in $\T U$ (and in $\T U_J$). In $\mu$-prolongations,
instead, the different vector components of $\T U$ (and of $\T
U_J$) are ``mixed'' by the prolongation operation.

\medskip\noindent
{\bf Remark 15.} It is known that $\mu$-symmetries (and hence
$\la$-symmetries, which are a special case of the latter) are
related to \emph{nonlocal} symmetries; we will not discuss this
relation here \cite{CF,MuRom5,MuRom12}.

\subsubsection{ODEs}

In the case of ODEs one just replaces the scalar function $\la :
J^1 M \to \R$ with a \emph{matrix} function $\La : J^1 M \to
\mathtt{Mat} (n)$ (more generally, $\La : J^1 M \to \T U$) and
define a ``$\La$-prolongation'' (which is just a special case of
$\mu$-prolongation, for $\mu = \La \d x$)
\begin{eqnarray} \psi^a_{(k+1)} &=& D_x
\psi^a_{(k)} \ - \ u^a_{(k+1)} \ D_x \, \xi \ + \ \La^a_{\ b} \,
\( \psi^b_{(k)} \ - \ u^b_{(k)} \, \xi \) \nonumber \\ &=& (D_x \, I \, + \,
\La )^a_{\ b} \, \psi^b_{(k)} \ - \ u^b_{(k+1)} \ (D_x \, I \, +
\, \La )^a_{\ b} \, \xi \ . \label{eq:Laprol} \end{eqnarray}

In this ODE case we just have $\mu = \La \, \d x $
(only one component), and \eqref{eq:MCH} is identically satisfied.

\medskip\noindent
{\bf Remark 16.} The IBDP property is in general not holding for
$\mu$-prolonged vector fields, not even in the ODEs framework; the
exception is the case where the $\La_i$ are diagonal matrices.

\subsubsection{Recursion formula}

The $\mu$-prolongation $X^{(k)}_\mu$, which we will now write in
components as $X^{(k)}_\mu = \xi^i \pa_i + (\psi^a_J)_{(\mu)}
\pa_a^J $, of a vector field $X$ in $M$ is defined through
\eqref{eq:muprol}; however in some cases and applications it is
relevant to characterize these in terms of the difference \beq
F^a_J \ := \ \( \psi^a_J \)_\mu \ - \ \( \psi^a_J \)_0 \ . \eeq It
can be shown \cite{CGMor,GMor} that the $F^a_J$ satisfy the
recursion relation \beql{eq:Fmu} F^a_{J,i} \ = \ \de^a_b \ \[ D_i
\, \( \Ga^J \)^b_c \] \ (D_i Q^c )  \ + \ \( \La_i \)^a_b \ \[
\(\Ga^J \)^b_c \ (D_J Q^c ) \ + \ D_j Q^b \]   \ , \eeq where we
have written \beql{eq:Q} Q^a \ = \ \vphi^a \ - \ u^a_i \, \xi^i \
, \eeq and the $\Ga^J$ are certain matrices whose detailed
expression can be computed \cite{CGMor,GMor} but is not essential.

\medskip\noindent
{\bf Remark 17.} With the notation \eqref{eq:Q}, the set $I_X$ of
$X$-invariant functions is characterized by $Q^a |_{I_X} = 0$. It
follows from \eqref{eq:Fmu} that $X^{(k)}_\mu$ coincides with
$X^{(k)}_0$ on $I_X$.

\section{Collective twisted symmetries: $\s$-symmetries}
\label{sec:sigma}

Let us consider the vector structure in $\T U$ and more generally
in $\T^k U$. We have seen that in $\la$-prolongations different
components (in terms of this structure) of a vector field ``do not
mix'' under the prolongation operation, while in
$\mu$-prolongations they do indeed ``mix''.

As mentioned above, Pucci and Saccomandi \cite{PuS} observed that
(in the scalar case) $\la$-prolongations are the \emph{only}
vector fields in $J^k M$ which have the same characteristics as
\emph{some} standardly prolonged vector field.

One can extend this approach to \emph{distributions} generated by
sets -- in involution \emph{\`a la Frobenius} -- of standardly
prolonged vector fields, and wonder if there is some deformation
of the prolongation operations such that a set of vector fields
obtained by this generate the same distribution in $J^k M$ as some
other set of vector fields prolonged in the standard way.

This problem was tackled by Cicogna \emph{et al.} in a series of
papers \cite{CGWs1,CGWs2,CGWs3,CGWs4} and the answer is that the
most general class of systems with this property is provided by so
called \emph{$\s$-prolonged} sets of vector fields\footnote{This
refers to the $(r \times r)$ matrix $\s$, where $r$ is the
cardinality of the set of vector fields, which appears in the
$\s$-prolongation formula \eqref{eq:sprol}, see below.}. Note that
here the deformation of the prolongation operation involves
\emph{sets} (more precisely, an involutive system) of vector
fields, and not a single one. We also stress that we are working
in the frame of ODEs, hence with only one independent variable
$x$.\footnote{Actually $\s$-prolongations and symmetries can also
be defined in the framework of PDEs (they go then under the name
of $\chi$-symmetries), but here we are mainly concerned with
Dynamical Systems.}

Given vector fields $X_\a$ ($\a = 1 , ... , r $) in $M$, written
in local coordinates as \beq X_\a \ =  \ \xi_\a \, \pa_x \ + \
\vphi^a_\a \, \pa_a \ , \eeq and satisfying the Frobenius
involution relations \beql{eq:frob} \[ X_\a , X_\b \] \ = \ f_{\a
\b}^\ga \ X_\ga \eeq with $f_{\a \b}^\ga : M \to \R$ smooth
functions on $M$, the $\s$-prolonged vector fields $Y_\a$ on $J^k
M$ are written as \beq Y_\a \ = \ \xi_\a \, \pa_x \ + \ \(
\psi^a_k \)_\a \, \pa_a^k \eeq where $(\psi^a_0)_\a = \vphi^a_\a$
and \beql{eq:sprol} \( \psi^a_{k+1} \)_\a \ = \ \( D_x (\psi^a_k
)_\a \ - \ u^a_{k+1} \ D_x \, \xi_\a \) \ + \ \s_\a^{\ \b} \ \(
(\psi^a_k)_\b \ - \ u^a_{k+1} \, \xi_\b \) \ . \eeq

\medskip\noindent
{\bf Lemma 6.} Let $X_\a$ satisfy \eqref{eq:frob}, and
\emph{assume} their $\s$-prolongations $Y_\a$ are in involution.
Then the set $\{ Y_\a \}$ has the IBDP property\footnote{We
specify that in this case the IBDP property should be meant as
follows: if $\eta$ and $\zeta_{(k)}$ are independent common
differential invariants for all of the $Y_\a$, then so are the
$\zeta_{(k+1)} := (D_x \zeta_{(k)})/(D_x \eta)$.}.

\medskip\noindent
{\bf Remark 18.} For fields $X_\a$ satisfying \eqref{eq:frob} and
$Y_\a$ their $\s$-prolongations, in general, \beq \[ Y_\a , Y_\b
\] \ \not= \ f_{\a \b}^\ga \ Y_\ga \ . \eeq However, the $Y_\a$
can happen to be still in involution, or to be embedded in  set of
vector fields in involution of non-maximal rank. This is why in
Lemma 6 the involution property of the $Y_\a$ has to be assumed.

\medskip\noindent
{\bf Remark 19.} While the $\mu$-prolongations mix different
vector components of the same vector field, here corresponding
components of different vector fields are mixed. For $r=1$ we are
back to the case of $\la$-prolongations.

\medskip\noindent
{\bf Remark 20.} If $\s = \mathtt{diag} (\la_1 , ... , \la_n )$ is
a diagonal matrix (but not a multiple of the identity) we have
different vector fields undergoing $\la$-prolongations with
different functions $\la_i$. In the case of $\s = \la I$ we are
back to the case of $\la$-prolongations (in general, applied to a
set of vector fields in multidimensional space).

\section{The use of twisted symmetries}
\label{sec:use}

We have so far discussed the definition of different types of
twisted prolongations and hence of twisted symmetries. We would
now like to discuss how these are applied in the study of
differential equations. In doing this one should distinguish
between ODEs and PDEs, recalling that -- as also stressed above --
the very aim of symmetry theory is different in these two
contexts.

We will always assume that the vector field $X$ is a twisted
symmetry (of different types) of the equations under study.

\subsection{The use of $\la$-symmetries}

If $X$ is a symmetry for an equation $\Delta$ of order $n$, this
means that $\Delta$ can be written in terms of the differential
invariants for $X^{(n)}_\mu$. On the other hand, as we have seen
above (Lemma 5), $\la$-prolonged vector fields enjoy the IBDP.
This implies that passing to $\la$-symmetry-adapted coordinates,
one can indeed rewrite the equation in terms of differential
invariants of order zero and one and their total derivatives,
implementing the reduction procedure sketched in Section
\ref{sec:sode}.

In other words, the usual symmetry reduction algorithm can be
applied also in the case of $\la$-symmetries, which are as useful
as standard ones in the study of ODEs.\footnote{We mention in
passing that $\la$-symmetries have also been used for the
reduction of \emph{discrete} equations \cite{LNR12,LR10}; this
lies outside our scope here.}

\subsection{The use of $\mu$-symmetries}

As mentioned above, $\mu$-symmetries were intended for application
in the study of PDEs. Here the key fact is \eqref{eq:Fmu} (see
also Remark 17); in fact, in studying PDEs by the symmetry
approach one is aiming at determining invariant solutions, and
\eqref{eq:Fmu} shows that when restricting to $X$-invariant
solutions it makes no difference to consider standard
prolongations or $\mu$-prolongations. This also entails that we
can use the same methods and techniques familiar from the case $X$
is a standard symmetry also in the case $X$ is a $\mu$-symmetry
(and in general not a proper symmetry).

\medskip\noindent
{\bf Remark 21.} This also shows that $\mu$-symmetries of a given
equation are strong candidates for being also (standard)
conditional symmetries \cite{LeWin,PuSweak}, or partial symmetries
\cite{CGpart}, for the same equation.

\medskip\noindent
{\bf Remark 22.} The situation is different in the case of ODEs
(this case was studied by Cicogna (he speaks in this case of
$\rho$-symmetries, the $\rho$ standing for ``reducing'', see
below) \cite{Cds1,Cds2}). In this case one can proceed pretty much
as in the standard reduction procedure up to a (relevant) feature:
that is, the reconstruction equation, which in the standard case
amounts to a quadrature, is now a proper differential equation,
and its solution may very well be very hard, or turn out to be
impossible. See \cite{Cds1,Cds2} for details.

\subsection{The use of $\s$-symmetries}

It follows immediately by Lemma 6 that $\s$-symmetries can also be
used to reduce (systems of) ODEs in the same way and with the same
procedure as in the standard case. Once again, the key fact is
that this standard reduction procedure \cite{\symmref} is actually
based on the IBDP.

It should be stressed, however, that in this case there is a
further condition to be satisfied, i.e. that the $Y_\a$ are (or
can be completed in a nontrivial way -- that is, without spanning
the whole tangent space -- to a system) in involution.

\medskip\noindent
{\bf Remark 23.} In this context, it should also be mentioned that
determining $\s$-symmetries is in general a nontrivial task
(recall that the determination of standard symmetries is often
computationally demanding, but always algorithmic); but when one
considers as differential equation a perturbation of a system for
which symmetries are known, $\s$-symmetries can be sought for as
deformation of the symmetries for the unperturbed system; see
\cite{CGWs3} (and Section \ref{sec:pert} below) for details.

\section{Twisted symmetries and gauge transformations}
\label{sec:gauge}

It appears that twisted symmetries are related to gauge
transformations, and indeed the operators $\nabla_i = D_i + \La_i$
appearing in $\mu$-prolongations look very much like a covariant
derivative. We will now make this relation more precise. For this,
it is convenient to consider just vertical vector fields,
including evolutionary representatives of general vector fields in
$M$.

\medskip\noindent
{\bf Lemma 7.} Let $X = Q^a \pa_a $ and $\wt{X} = \wt{Q}^a \pa_a$
be vertical vector fields on $(M,\pi_0,B)$, $A : M \to
\mathtt{Mat}(\R,q)$ (with $q = \mathtt{dim} (U)$) a nowhere zero
smooth matrix function, and $\wt{Q}^a = A^a_{\ b} Q^b$. Then
\beql{eq:mugauge} A \ \( X^{(n)}_\mu \) \ = \ \wt{X}^{(n)}_0 \ ; \
\ \ \mu \ = \ (D A) \, A^{-1} \ . \eeq

\medskip\noindent
{\bf Remark 24.} The relation between $X^{(n)}_\mu$ and
$\wt{X}^{(n)}_0$ in \eqref{eq:mugauge} should be meant as follows:
if $X^{(n)}_\mu = \psi^a_J \pa_a^J$, with $\psi^a_0 = Q^a$ and the
$\psi^a_J$ for $1 \le |J| \le n$ obtained by the
$\mu$-prolongation formula, and $\wt{X}^{(n)} = \wt{\psi}^a_J$
with $\wt{\psi}^a_0 = \wt{Q}^a$ and the $\wt{\psi}^a_J$ for $1 \le
|J| \le n$ obtained by the standard prolongation formula, then the
relation \beq \wt{\psi}^a_J \ = \ A^a_{\ b} \, \psi^b_J \eeq holds
for any $a$ and $J$, $0 \le |J| \le n$.

\medskip\noindent
{\bf Remark 25.} The relations stated by Lemma 7 can be encoded in
a commutative diagram:
\beq \begin{matrix} X & \mapright{A} & \wt{X} \\
\mapdown{\mu-prol} & & \mapdown{prol} \\
X^{(n)}_\mu & \mapright{A} & \wt{X}^{(n)}_0 \end{matrix} \eeq
where $A$ and $\mu$ are related by $\mu = (D A)  A^{-1}$.

\medskip\noindent
{\bf Remark 26.} Lemma 7 is \emph{not} stating that any
$\mu$-prolonged vector field is obtained as the gauge transformed
of a standardly prolonged one; this relation only holds for
vertical vector fields. If $X = Q^a \pa_a$ is the evolutionary
representative of a generic vector field $X_g = \xi^i \pa_i +
\vphi^a \pa_a$, hence $Q^a = \vphi^a - u^a_i \xi^i$, its
components $Q^a$ satisfy the relations $(\pa Q^a / \pa u^b_i ) = -
\delta^a_b \xi^i$. These are obviously not satisfied in general by
the components $\wt{Q}^a = A^a_{\ b} Q^b$ of $\wt{X}$ (now $(\pa
\wt{Q}^a / \pa u^b_i ) = - A^a_b \xi^i$), hence we cannot
interpret the gauge-transformed vector field as the evolutionary
representative of a vector field in $M$ \cite{Gtwist2}.

\medskip\noindent
{\bf Remark 27.} The previous Remark also means that the
connection between $\mu$-prolongations and gauge transformations
is only transparent when we consider the action of vector fields
on $\Sigma (M)$, the set of sections on $M$. It also explains why
we can have twisted symmetries for equations having no standard
symmetries.

\medskip\noindent
{\bf Remark 28.} More details on the interrelations between
(different kinds of) twisted symmetries and gauge transformations
are provided e.g. in \cite{Ggauge1,Ggauge2,Ggauge3}.

\section{Twisted symmetries and Frobenius theory}

The existence of a relation between twisted symmetries and gauge
transformations was more and less evident since the introduction
of $\la$-symmetries by Muriel and Romero, and so the
generalization of $\la$-symmetries to $\mu$-symmetries was, in
this sense, not surprising.

It was much less obvious that symmetries and twisted symmetries
could be generalized in a different direction, focusing on
\emph{sets} (actually, involutive systems) of vector fields rather
than on single ones\footnote{One speaks therefore of
``collective'' twisted symmetries. Actually here we will only deal
with the case of ODEs and Dynamical Systems ($\s$-symmetries),
rather than general PDEs ($\chi$-symmetries). For the latter, the
interested reader is referred to \cite{CGWs3,Gtwist2}.}. As
already mentioned, the key step in this direction was provided by
Pucci and Saccomandi \cite{PuS}, who stressed the symmetry
relation involves the integral lines of (prolongations of)
symmetry vector fields, not the way in which the flow defined by
the vector field travels on these.

The geometrical idea behind $\s$-symmetries is that (standard)
symmetry vector fields for the equation $\De$ of order $n$ are the
vector fields on $M$ whose (standard) prolongation to $J^n M$
belongs to the \emph{distribution} tangent to $S_\De$ in $\T J^n
M$. Focusing on the distribution -- rather on the single vector
fields, i.e. the ``usual'' generators of the distribution -- has
an obvious consequence: we can change the generators of the
distribution.

In particular, if we are dealing with ODEs, we would like to be
able to change the generators of the distribution (to have more
freedom), but at the same time be sure that the key tool for ODE
reduction, i.e. the IBDP, is still at work.

The idea of $\s$ symmetries is exactly this: the $\s$-prolongation
is the most general way of twisting prolongation by mixing
\emph{different} vector fields in such a way that the IBDP still
holds, and hence so that twisted symmetries can be of use for the
reduction of ODEs.

\medskip\noindent
{\bf Lemma 8.} Let $\mathcal{X} = \{ X_1 , ... , X_r \}$ be a set
of vector fields on $M$; and let the vector fields $\mathcal{Y} =
\{ Y_1 , ... , Y_r \}$ on $J^n M$ be their $\s$-prolongation.
Consider also $A : M \to \mathtt{Mat} (\R,q)$ (where $q =
\mathtt{dim} (U)$) a nowhere singular matrix function on $M$, such
that $\s = A^{-1} (D_x A)$; and the set $\mathcal{W} = \{ W_1 ,
... , W_r \}$ of vector fields on $M$ given by $W_\a = A_\a^{\ \b}
X_\b$, with $\mathcal{Z} = \{ Z_1 , ... , Z_r \}$ on $J^n M$ their
standard prolongation. Then, $Z_\a = A_\a^{\ \b} Y_\b$.


\medskip\noindent
{\bf Remark 29.} The relations stated by Lemma 8 can be encoded in
a commutative diagram:
\beq \begin{matrix} \{X_i \} & \mapright{A} & \{ W_i \} \\
\mapdown{\s-prol} & & \mapdown{prol} \\
\{ Y_i \} & \mapright{A} & \{ Z_i \} \end{matrix} \eeq
where $A$ and $\s$ are related by $\s = A^{-1} (D A)$.

\medskip\noindent
{\bf Remark 30.} Traditionally, in the symmetry analysis of
differential equations one focuses on the Lie algebra structure of
symmetry vector fields. Passing to consider Frobenius reduction
means one is instead focusing on the \emph{Lie module}
structure.\footnote{Here we mean a module over the algebra
$C^\infty (M,R)$ of (smooth) real functions on $M$. Note that
while some equations (in particular all equations which are linear
or can be linearized by a change of variables) have an infinite
dimensional Lie algebra of symmetries, their set of symmetries is
always finitely generated as a Lie module.}

\medskip\noindent
{\bf Remark 31.} The determination of standard symmetries is
algorithmic for equations of order $n \ge 2$, but is especially
difficult for first order equations -- even more so for first
order ODEs, i.e. Dynamical Systems -- albeit in general we always
have infinitely many symmetries in this case. In the case of
Dynamical Systems, an interesting possibility was noted by Cicogna
\cite{CGWs3}: if we consider the perturbations of a symmetric
Dynamical Systems, $\s$-symmetries can be looked for by building
$\s$ as a perturbation to the identity matrix. See Section
\ref{sec:pert} below.

\medskip\noindent
{\bf Remark 32.} The possibility of using sequentially different
symmetries for Frobenius reduction rests -- like in the case of
standard reduction -- on a suitable involution structure; that is,
we should have a solvable Lie-module.

\medskip\noindent
{\bf Remark 33.} More details on $\s$-prolongations and
symmetries, including their geometrical meaning, is provided in
the papers \cite{CGWs1,CGWs2,CGWs3,CGWs4}; see also
\cite{Gtwist2}.

\section{Twisted symmetries and variational problems}
\label{sec:variational}

The theory of twisted symmetries was developed mainly referring to
ODEs and evolution PDEs. But we know that in Physics a special
role is played by problems admitting a \emph{variational
formulation}. In this case, the relation between symmetries and
conservation laws is embodied by the classical Noether theorem
\cite{Arn,YKS,Olv1}. It is thus natural to wonder if there is a
version of Noether theorem applying to \emph{twisted} symmetries.

Only partial results exist in this direction; these are concerned
with $\la$-symmetries \cite{MRO,RMO} and with $\mu$-symmetries
\cite{CGNoether}, while it seems no result is available dealing
with $\s$-symmetries.

\subsection{Variational problems and $\la$-symmetries}

The relation between $\la$-sym\-me\-tri\-es and Euler-Lagrange
equations has been considered in a by now classical work of
Muriel, Romero and Olver \cite{MRO}; lately new results in this
direction have been obtained by Ruiz, Muriel and Olver \cite{RMO}.

\subsubsection{Single $\la$-symmetry of a variational problem}

We consider variational problems defined by a Lagrangian density
$L$, hence by \beql{eq:lag} S [u] \ = \ \int L (x,u_{[n]}) \, d x
\ ; \eeq here $x \in \R$, $u \in \R$. To this problem are
associated the Euler-Lagrange equations \beq E[L;u] \ = \
\sum_{k=0}^n \( - \, D_x \)^k \ \( \frac{\pa L}{\pa u_k} \) \ = \
0 \ . \eeq

A vector field $X = \xi (x,u) \pa_x + \vphi (x,u) \pa_u$ is a
standard variational symmetry \cite{Olv1} if there is a function
$F : J^n M \to \R$ such that \beq X^{(n)} ( L) \ + \ L \ (D_x \xi
) \ = \ D_x F \ . \eeq This definition is generalized by saying
that $X$ is a \emph{variational $\la$-symmetry} if there is a
function $F : J^n M \to \R$ such that \beq X^{(n)}_{(\la)} ( L) \
+ \ L \ [(D_x + \la) \xi ) \ = \ (D_x +\la ) F \ . \eeq If $X$ is
a variational $\la$ symmetry for $L$, it is such also for $\wt{L}
= L + (D_x f)$, for any $f : J^n M \to \R$, i.e. for any
equivalent Lagrangian \cite{MRO}.

Variational $\la$-symmetries lead to reduction of order for the
variational problem in the same way as standard variational
symmetries. More precisely, Muriel, Romero and Olver prove the
following result (Theorem 1 in \cite{MRO}).

\medskip\noindent
{\bf Lemma 9.} Let $S[u]$ as in \eqref{eq:lag} be an
n$^{th}$-order variational problem with Euler-Lagrange equation
$E[L;u] = 0$ of order $2n$. Let $X$ be a variational
$\la$-symmetry, where $\la : J^1 M \to \R$ is smooth. Then there
exists a variational problem \beq \^S [w] \ = \ \^L
(\wt{x},w_{[n-1]}) d \wt{x} \eeq of order $n - 1$, with
Euler-Lagrange equation $E[\^L;w] = 0$ of order $2n-2$, such that
a $(2n-1)$-parameter family of solutions of $E[L;u] = 0$ can be
found by solving a first-order equation from the solutions of the
Euler-Lagrange reduced equation $E[\^L;w] = 0$.
\bigskip

As for the Noether theorem, this essentially follows (for standard
symmetries) from \beql{eq:noe} X^{(n)} (L) \ = \ Q \ E[L] \ + \
D_x F \eeq where $F$ is some function $F: J^n M \to \R$, and $Q =
\vphi - u_x \xi$ is the characteristic of the (evolutionary
representative of) $X$.

In the case of $\la$-prolongations, one can prove \cite{MRO} that
there is some $F$ such that \beql{eq:noela} X^{(n)}_\la (L) \ = \
Q \ E[L] \ + \ (D_x \, + \, \la ) F \ . \eeq Then the following
result (which is Theorem 2 in \cite{MRO}) follows.

\medskip\noindent
{\bf Lemma 10.} Let $X$ be a variational $\la$-symmetry of the
variational problem \eqref{eq:lag}, and $Q$ the characteristic of
$X$. Then there exists $P[u] : J^n M \to \R$ such that \beq Q \
E[L] \ = \ (D_x + \la ) P \ . \eeq

\medskip\noindent
{\bf Remark 34.} While standard variational symmetries of a
variational problem are symmetries of the corresponding
Euler-Lagrange equations, $\la$-symmetries of the variational
problem are in general \emph{conditional} symmetries of the
Euler-Lagrange equations \cite{MRO}  (see also Remark 17 in this
respect).

\medskip\noindent
{\bf Remark 35.} If $X$ is a variational $\la$-symmetry of
\eqref{eq:lag}, and $P[u]$ is the functional given in Lemma 10,
then $X$ is a $\la$-symmetry of the equation $P[u] = 0$. The
reduction of this equation through $X$ is (up to multipliers) the
reduced equation of the Euler–Lagrange equation corresponding to
$X$, according to Lemma 9 \cite{MRO}.

\subsubsection{Multiple $\la$-symmetry of a variational problem}

In more recent work, Ruiz, Muriel and Olver \cite{RMO} studied
variational problems systems which admit several $\la$-symmetries
$X_i$, where for each of them a \emph{different} function $\la_i$
defines $\la$-prolongation.

They considered in particular the case of two such
$\la$-symmetries $\{ X, Y \}$, subject to the ``solvability
condition'' \beql{eq:solvla} \[ X^{(n)}_{\la_1} , Y^{(n)}_{\la_2}
\] \ = \ h \, X^{(n)}_{\la_1} \ . \eeq

Then the $\la$-symmetries can be used (in the proper order!) to
perform \emph{two} symmetry reductions of the variational
problem.\footnote{It appears that the result can be extended to
$k$ $\la_i$-symmetries with a suitable solvability condition, i.e.
generating a solvable Lie module.} In particular, one can prove
\cite{RMO} that:

\medskip\noindent
{\bf Lemma 11.} Let \eqref{eq:lag} be an $n$-th order variational
problem with Euler-Lagrange equation $E[L;u] = 0$ of order $2n$.
Let $(X_1, \la_1), (X_2, \la_2)$ be variational $\la$-symmetries
that form a solvable pair, as in \eqref{eq:solvla}. Then there
exists a variational problem $\^S = \int \^L [x,z_{n-2}] d x$ of
order $n-2$ such that a $(2n-2)$-parameter family of solutions of
$E[L;u] = 0$ can be reconstructed from the solutions of the
associated $(2n-4)$-th order Euler-Lagrange equation $E[\^L;z]
 = 0$ by solving two successive first order ordinary differential equations.

\medskip\noindent
{\bf Remark 36.} Albeit $\la$-prolongations with different
functions $\la_i$ for different vector fields fit within
$\s$-prolongations  -- see in particular Remark 20 -- it should be
stressed that variational problems have never been studied in that
framework. Thus Lemma 11 calls for a full study of Frobenius
reduction in the variational context.

\subsection{Variational problems and $\mu$-symmetries}

A different approach to twisted symmetries (in particular,
$\mu$-symmetries) in variational problems was considered by
Cicogna {\it et al.} \cite{CGNoether}. In this work they show
that $\mu$-symmetries are associated to so called
$\mu$-conservation laws\footnote{A conservation law is a relation
of the type $D_i \cdot {\bf P}^i = 0$ for some vector ${\bf P}$; a
$\mu$-conservation law reads $\mathtt{Tr} \( \nabla_i \cdot {\bf
P^i} \) = 0$, with $\nabla_i = D_i +\Lambda_i$.} and in the end,
for variational problems with a single independent variable
(dynamical variational problems) and $\La = \la I$, to
conditionally conserved quantities
\cite{LC,PuRos,PuRos2,SarCan,SarCan2}.

These are quantities such that only \emph{some} of their level
sets correspond to invariant manifolds -- while for proper
conserved quantities \emph{all} the level sets are dynamically
invariant; note that the result is strictly related to the one
mentioned in Remark 34.

There is a different way of looking at variational problems, in
particular dynamical variational problems, with $\mu$-symmetries;
this descends from the relation between $\mu$-prolongations and
gauge transformations.

If we think of such a problem as arising -- through a gauge
transformation -- from one in which the $\mu$-prolonged vector
field was prolonged in the standard way, i.e. perform the needed
inverse gauge transformation (see section \ref{sec:gauge}), we
should think that the original variational problem was a different
one, and Euler-Lagrange equations were also different. In
particular, the role of $D_i$ in the Euler-Lagrange operator would
be taken by $\nabla_i = D_i + \La_i$. Thus, e.g., in the case of
Mechanics the $\mu$-Euler-Lagrange equations read \beq \frac{d}{d
t} \frac{\pa L}{\pa {\dot q}^i} \ - \ \frac{\pa L}{\pa q^i} \ = \
\( \La^T \)_i^{\ j} \, \frac{\pa L}{\pa {\dot q}^j} \ . \eeq One
can then show that (for further details, proofs and examples, see
\cite{CGNoether}):

\medskip\noindent
{\bf Lemma 12.} If $L$ is a first-order Lagrangian admitting the
vector field $X = \vphi^a \pa_a$ as a $\mu$-symmetry, then ${\bf
P}$ of components $P^i = \vphi^a \pi^i_a$ (where $\pi^i_a = \pa L
/ \pa u^a_i$) defines a \emph{standard} conservation law, $D_i P^i
= 0$, for the flow of the associated $\mu$-Euler-Lagrange
equations.

\section{Twisted symmetries and perturbations of Dynamical Systems}
\label{sec:pert}

As mentioned above (see Remarks 23 and 31), $\s$-symmetries turn
out to be specially suited for the investigation of perturbations
of symmetric Dynamical Systems. It should be stressed that in this
case also the determination of $\s$-symmetries (which is, as for
all types of twisted symmetries, a non-algorithmic task) turns out
to be facilitated.

The type of result one can obtain in this direction is illustrated
by  the following result (Theorem 4 in \cite{CGWs3}):

\medskip\noindent
{\bf Lemma 13.} Let the dynamical system \beq \frac{d x^i}{d t} \
= \ f^i (x) \eeq admit the vector fields $X_\a = \vphi^i_\a \pa_i
$ as standard symmetries, and let these span a Lie
algebra\footnote{The real constants $c_{\a \b}^\ga$ being the
structure constants.}, \beq
\[ X_\a , X_\b \] \ = \ c_{\a \b}^\ga \, X_\ga \ . \eeq Consider
moreover the vector fields $Y_\a$ in $J M$ obtained as
$\s$-prolongations of the $X_\a$ with \beql{eq:sL13} \s_\a^{\ \b}
\ = \ c_{\a \ga}^\b \, F^\ga \ + \ X_\a (F^\b) \ . \eeq Then:
$(i)$ The $Y_\a$ are in involution and satisfy the same
commutation relations as the $X_\a$; $(ii)$ any dynamical system
of the form \beql{eq:pds} \frac{d x^i}{d t } \ = \ f^i (x) \ + \
\sum_{\a=1}^r F^\a (x) \, \vphi^i_\a (x) \eeq admits the set of
$X_\a$ as $\s$-symmetries -- with $\s$ given by \eqref{eq:sL13} --
and hence can be reduced via these.

\medskip\noindent
{\bf Remark 37.} The form \eqref{eq:pds} of systems which can be
dealt with in this way may seem too specific, but it includes at
least one relevant class, i.e. that of systems in Poincar\'e-Dulac
\emph{normal form} \cite{ArnGM,CGbook}. In fact, let $f(x) = A x$
with $A$ a semi-simple matrix; then we consider $\vphi_\a (x) =
B_\a x$ with matrices such that $[A,B_\a] = 0$ (these obviously
form a Lie algebra $\mathcal{G}$, and $B_0 = A$ is always in the
set). Then the polynomial vector fields which admit $X_0 = (A x)^i
\pa_i $ as symmetry are just those written in the form
\eqref{eq:pds}, with $F^\a (x)$ generators for the ring of
$X_0$-invariant functions. See \cite{CGWs3} for details and
examples, as well as \cite{GGSW,GWNF,GWNF2} for related matters.

\medskip\noindent
{\bf Remark 38.} Orbital reduction of dynamical systems
\cite{HaWa,Wal99,Wal99b} can be dealt with in a similar manner; we
will not discuss this here.

\section{Conclusions}

The theory of twisted symmetries of differential equations has
been created in 2001; it passed from a smart observation by its
creators \cite{MuRom1,MuRom2} to a coherent set of results, and
from an analytic formulation to a geometrical one. In particular,
in the course of this travel several relations with \emph{gauge
transformations} and with the Frobenius theory of vector fields
have been uncovered, and it has been realized how twisted
symmetries become rather natural if one looks not at the standard
theory of (Lie) reduction, but at \emph{Lie-Frobenius reduction}
for differential equations.

It has also been realized that twisted symmetries -- in the form
of ``perturbed prolongations'' -- can be used to study
perturbations of symmetric equations and in particular symmetric
Dynamical Systems; this part of the theory definitely awaits
further developments.

Similarly, the study of twisted symmetries (and their use) for
variational problems is in its initial phase, and is worth
receiving further attention.

Albeit we have not touched this topic at all, I would like to
mention also that in the recent wave of interest for symmetries of
\emph{stochastic} differential equations (see \cite{GGPR} and
references therein)  there is not yet any work studying the role
(if any) of twisted symmetries in that context.

I hope these pages can help attracting  mathematicians to this
nice and promising field; it would be even nicer if our young
friend Juergen Scheurle himself could contribute to the topic.

\section*{Acknowledgements} Work partially supported by SMRI and GNFM-INdAM.


\end{document}